\documentclass[letterpaper,english,aps, prl, reprint, superscriptaddress]{revtex4-2}

\usepackage[T1]{fontenc}
\usepackage[latin9]{inputenc}
\usepackage{amsmath}
\usepackage{amssymb}
\usepackage{bbold}
\usepackage{graphicx}

\usepackage{times}
\usepackage{babel}

\PassOptionsToPackage{position=top}{subfig}
\usepackage{hyperref}
\hypersetup{
	colorlinks = true,
	allcolors = blue
}

\makeatletter   

\begin{document}

\title{Topological Anderson insulator and reentrant topological transitions in a mosaic trimer lattice}

\author{Xiatao Wang}
\affiliation{Institute of Theoretical Physics, State Key Laboratory of Quantum Optics Technologies and Devices, Collaborative Innovation Center of Extreme Optics, Shanxi University, Taiyuan 030006, China}

\author{Li Wang}
\email{liwangiphy@sxu.edu.cn}
\affiliation{Institute of Theoretical Physics, State Key Laboratory of Quantum Optics Technologies and Devices, Collaborative Innovation Center of Extreme Optics, Shanxi University, Taiyuan 030006, China}

\author{Shu Chen}
\affiliation{Beijing National Laboratory for Condensed Matter Physics, Institute
of Physics, Chinese Academy of Sciences, Beijing 100190, China}
\affiliation{School of Physical Sciences, University of Chinese Academy of Sciences,
Beijing 100049, China }

\date{\today}

\begin{abstract}
We study the topological properties of a one-dimensional quasiperiodic-potential-modulated mosaic trimer lattice.
To begin with, we first investigate the topological properties of the model in the clean limit free of quasiperiodic disorder based on analytical derivation and numerical calculations of the Zak phase $Z$ and the polarization $P$. 
Two nontrivial topological phases corresponding to the $1/3$ filling and $2/3$ filling, respectively, are revealed.
Then we incorporate the mosaic modulation and investigate the influence of quasiperiodic disorder on the two existing topological phases.
Interestingly, it turns out that quasiperiodic disorder gives rise to multiple distinct effects for different fillings.
At $2/3$ filling, the topological phase is significantly enhanced by the quasiperiodic disorder and topological Anderson insulator emerges. Based on the calculations of polarization and energy gap, we explicitly present corresponding topological phase diagram in the $\lambda-J$ plane.
While for the $1/3$ filling case,  
the quasiperiodic disorder dramatically compresses the topological phase, and strikingly, further induces the emergence of reentrant topological phase transitions instead.
Furthermore, we verify the topological phase diagrams by computing the many-body ground state fidelity susceptibility for both the $1/3$ filling and $2/3$ filling cases.
Our work exemplifies the diverse roles of quasiperiodic disorder in the modulation of topological properties,  and will further inspire more research on the competitive and cooperative interplay between topological properties and quasiperiodic disorder.
\end{abstract}

\maketitle

\textcolor{blue}{\em Introduction}--Topological properties and topological phase transitions of various quantum matter constitute a cutting-edge focus in contemporary condensed matter physics~\cite{science.1133734, Wang2017,RevModPhys.83.1057, RevModPhys.82.3045,RevModPhys.88.021004, Weng2014,Liu2019,Pesin2012,PhysRevLett.127.136802,PhysRevB.107.035113,PhysRevB.109.035114,RevModPhys.91.015005,Goldman2016,Lee2018,RevModPhys.91.015006}.
As is well known, topological materials are a class of emerging quantum systems whose quantum states exhibit nontrivial topological properties. The core characteristic lies in that the properties of the system's wave function are determined by topological invariants through bulk-edge correspondence in a global level, rather than by local details of the system. Thus topological states of quantum matter are favorably immune to weak disorder or local defects, which makes them highly valued in the field of quantum computation research~\cite{Pachos_2012}. 
And it is commonly recognized that when the disorder strength is large, topological phases will eventually be destroyed and transition into Anderson insulator~\cite{Anderson1958pr}. 

However, the emergence of the concept of topological Anderson insulators (TAIs)~\cite{PhysRevLett.102.136806,PhysRevLett.103.196805,PhysRevB.84.035110, PhysRevB.80.165316, PhysRevB.85.035107,PhysRevB.104.L161118,PhysRevB.102.205425,PhysRevB.100.184202,PhysRevB.85.195125} in recent years has overturned the conventional understanding that disorder may merely result in the impairment of topological properties. Studies have shown that moderate disorder sometimes  not only does not disrupt the existing topological phases, but also induce the transition from topologically trivial phases to topologically non-trivial phases, thereby enhancing the topological properties of the system. 
Thus, the interplay between disorder and topology has drawn extensive attention from physicists.
And more TAIs and their extensions~\cite{PhysRevLett.126.146802,PhysRevB.103.085408,PhysRevB.110.075422,10.21468/SciPostPhys.18.6.208,PhysRevLett.129.043902, PhysRevB.108.L081110, PhysRevB.110.085157, PhysRevB.103.085307, PhysRevA.105.063327, PhysRevB.108.085306,Zhang2024,PhysRevA.111.022214,Zhang2020,Lin2022,PhysRevLett.101.246810,doi:10.1126/science.aat3406,stutzer2018photonic,
PhysRevLett.125.133603,PhysRevResearch.6.L042038} have been uncovered, such as high-order topological Anderson insulators~\cite{PhysRevLett.126.146802,PhysRevB.103.085408,PhysRevB.110.075422,10.21468/SciPostPhys.18.6.208}, $\mathbb{Z}_2$ topological Anderson insulators~\cite{PhysRevLett.129.043902}, topological Anderson amorphous insulators~\cite{PhysRevB.108.L081110}, topological inverse Anderson insulators~\cite{PhysRevB.110.085157}, etc.

Moreover, beyond the monotonous destructive or enhancing behavior, it is shown that disorder, including quasiperiodic disorder, can also induce the emergence of reentrant topological phase transitions~\cite{PhysRevB.85.140508,PhysRevB.88.060509,PhysRevA.111.022214,ZhanpengLu2025024204,PhysRevB.85.195304,PhysRevB.88.155428,
PhysRevB.95.195408,PhysRevB.96.245118,PhysRevB.109.L020203,PhysRevB.111.115153,PhysRevB.111.155149,9jjd-vbp1,PhysRevB.101.045133,
zjyw-ln2n}.
Reentrant topological phase transition~\cite{FUJII1985715} constitutes an intriguing phenomenon in the field of topological matter, which features a non-monotonic sequence of topological-trivial-topological phase transitions emerged in a system as the relevant parameter varies.
This further demonstrates the richness of the physical connotations underlying the interplay between disorder and topological properties.
Thus, more intriguing phenomena underlying the interplay between disorder and topology are yet to be explored.
For example, natural questions arise,  is it possible to demonstrate the above-mentioned multiple effects of disorder on topological properties  in a single system. While most studies have focused on single-band or SSH-like models~\cite{PhysRevB.100.205302,PhysRevA.106.013305,PhysRevA.111.022214,PhysRevB.108.195103,PhysRevB.102.205425,PhysRevB.109.195124} to explore the topological effects of quasiperiodic disorder, what new can emerge in one-dimensional multi-band systems?
This basically gives rise to the core idea of this work.

In this work, we propose a one-dimensional mosaic trimer lattice model modulated by a quasiperiodic potential and investigate its topological properties in detail.
First, we analytically tackle this model in its clean limit, and demonstrate two nontrivial topological phases corresponding to the $1/3$ filling and $2/3$ filling, respectively. This is confirmed by numerical calculations of the Zak phase $Z$ and polarization $P$. 
Then we continue by taking into account the quasiperiodic disorder modulations and explore the interplay between the existing topology and quasiperiodic disorder.
It is found that in such a single model, quasiperiodic disorder exerts a multifaceted influence on the topological properties.
For the topological phase at $2/3$ filling, the quasiperiodic disorder yields pronounced topological enhancement.
The region of the topological phase is significantly expanded, and thus giving rise to the topological Anderson insulator (TAI).
While for the topological phase at $1/3$ filling, the role of the quasiperiodic disorder modulations turns out to be more intriguing and rich.
It is shown that quasiperiodic disorder suppresses the topological phase remarkably upon its introduction. 
With the increase of the disorder strength, the region of the topological phase shrinks. 
However, further increase in disorder induces trivial phase inside the topological phase, giving rise to the interesting reentrant topological phase transition.
More interestingly, for certain disorder strength, increasing the value of inter-cell hopping $J$ can also induce reentrant topological phase transition.
Furthermore, by computing the many-body ground state fidelity susceptibility at different fillings, we confirmed the two topological phase diagrams in the $\lambda-J$ plane.

\begin{figure}[tp]
\includegraphics[width=7cm]{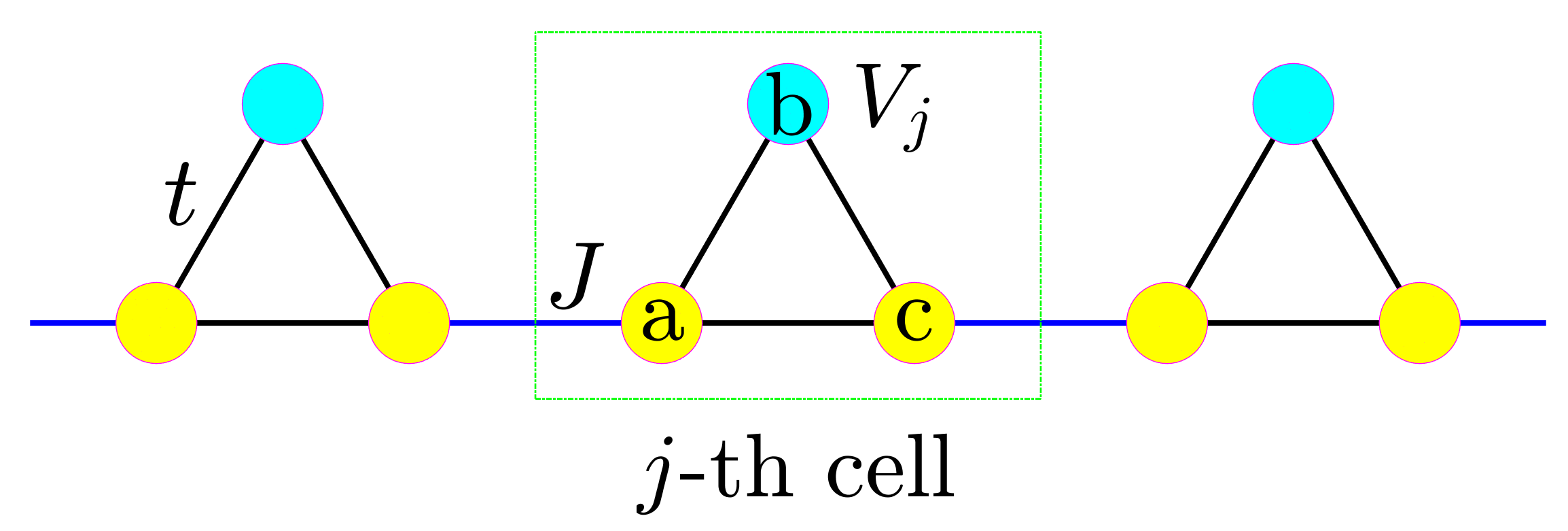}
\caption{Sketch of the one-dimensional mosaic trimer lattice.
Each small circle represents a lattice site. 
Black solid line represents intra-cell hopping \( t \), blue solid line denotes inter-cell hopping \( J \), and the quasi-periodic potential \( V_j \) exists only at \( b \) sites.  
}
\label{Fig01}
\end{figure}

\textcolor{blue}{\em The model and its clean limit}--We investigate the topological properties of a class of one-dimensional (1D) quasiperiodic-potential-modulated trimer lattice as depicted in Fig. \ref{Fig01}, which can be briefly described by the following tight-binding hamiltonian,
\begin{align}
H =& \sum_{j=1}^{N} \left[ t \left  (a_j^{\dagger} b_j +  b_j^{\dagger} c_j + c_j^{\dagger} a_j \right )+ J c_j^{\dagger} a_{j+1} +\text{H.c.} \right]  \nonumber \\
+ &\sum_{j=1}^{N} V_j b_j^{\dagger} b_j, \label{Htrimer}
\end{align}
where the quasiperiodic on-site potential solely on the $b$ sites reads $V_j = -\lambda \sec^2(2\pi \beta j + \theta)$.
 \( a_j^\dagger \), \( b_j^\dagger \) and \( c_j^\dagger \) (\( a_j \), \( b_j \) and \( c_j \)) are creation (annihilation) operators on the \( a \), \( b \) and \( c \) sublattices of the \( j \)-th unit cell, respectively. 
$N$ is the number of unit cells and $L = 3N$ is the total number of lattice sites. 
\( \lambda \) dictates the strength of the quasiperiodic potential. \( t\) and \( J \) correspond to the strength of intra- and inter-cell hopping, respectively (see Fig.~\ref{Fig01}). 
 \( \beta \) is an irrational number dictating the quasiperiodicity of the lattice. 
 \( \theta \) is a phase of the quasiperiodic potential. 
Throughout this work, we set \( \beta = (\sqrt{5}-1)/2\)  and \( \theta = 0 \) in the following calculations.

According to the time-independent Schr\"{o}dinger equation \( \hat{H}\vert \Psi \rangle= E \vert \Psi \rangle \), where \( \vert \Psi \rangle = \sum_{\alpha,j} \psi_{\alpha,j}  \alpha_j^{\dagger}  \vert 0 \rangle \)  with $\alpha$ denoting $a$, $b$, or $c$, one can obtain the eigenequations as following,
\begin{align}
E\psi_{a,j} =& J\psi_{c,j-1} + t (\psi_{b,j}+\psi_{c,j}),  \label{psia} \\
E\psi_{b,j} =& V_j\psi_{b,j} + t (\psi_{a,j} + \psi_{c,j}), \label{psib}\\
E\psi_{c,j} =& t (\psi_{a,j}+\psi_{b,j}) + J\psi_{a,j+1} .  \label{psic}
\end{align}

\begin{figure}[tbp]
\includegraphics[width=6.7cm]{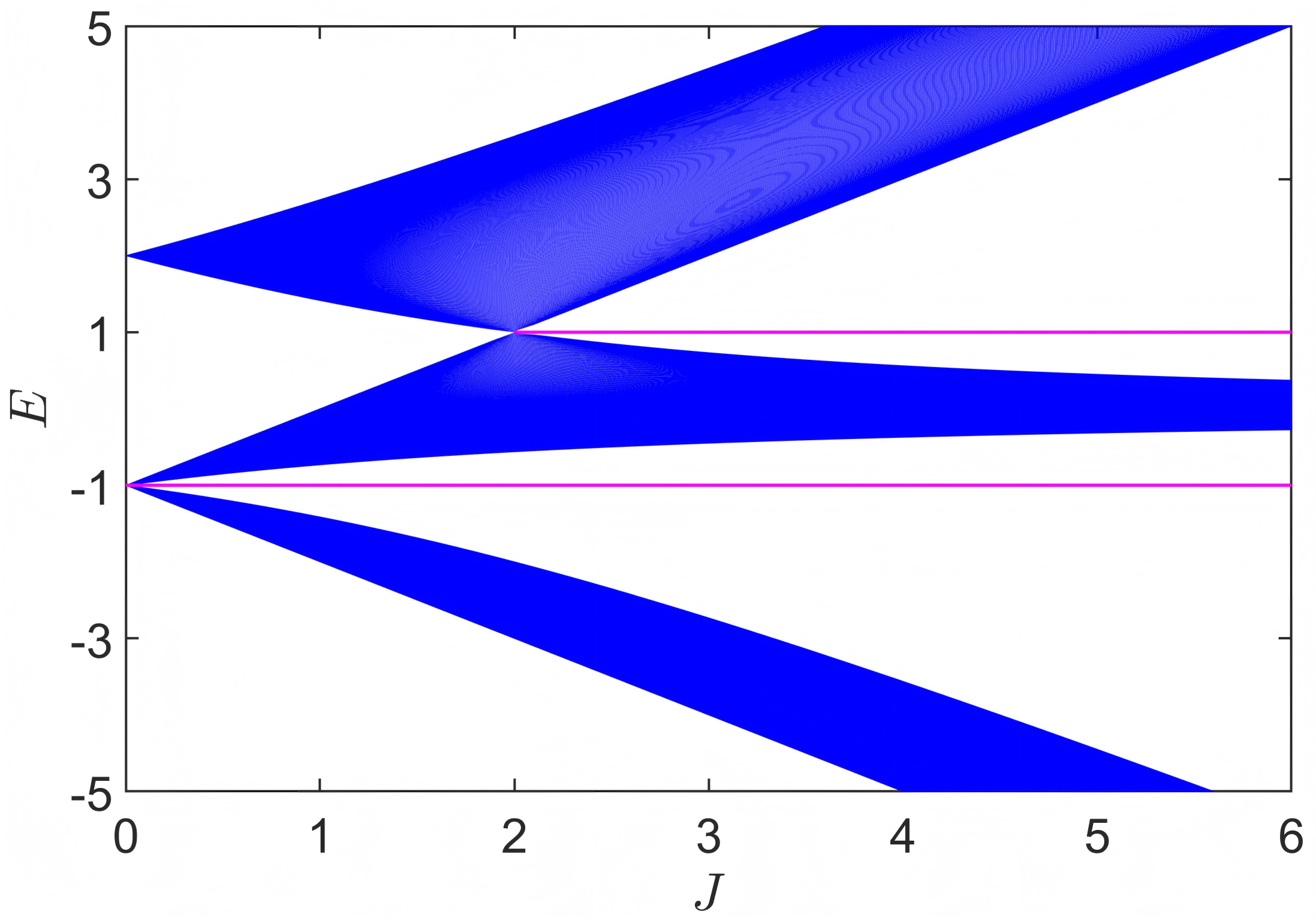}
\caption{
Energy spectra as a function of  the inter-cell hopping strength $J$ for the trimer lattice model in Eq.~(\ref{Htrimer}) with open boundary condition (OBC) in its clean limit. The topological edge states are highlighted with magenta color. 
Other parameters are $L=3N=699$, $t=1$.
}\label{Fig02}
\end{figure}

Before delving into the topological properties under the most general scenario with quasiperiodic modulation, we first examine the topological properties of the model for the simplest case, i.e., its clean limit with $\lambda=0$, which not only serves as a starting point for investigating the general case but also acts as a crucial benchmark, which enables the clear highlighting of the novel physical effects inherent in the general scenario through direct comparison.

In the clean limit with $\lambda=0$, Eq.~(\ref{psib}) reduces to $\psi_{b,j} = t (\psi_{a,j} + \psi_{c,j})/E$. 
Substituting it into Eqs.~(\ref{psia}) and (\ref{psic}), one then obtains,
\begin{align}
\tilde{E} \psi_{a,j} &=  T \psi_{c,j} + J \psi_{c,j-1},  \label{ssha} \\
\tilde{E} \psi_{c,j} &= T \psi_{a,j} + J \psi_{a,j+1}, \label{sshb}
\end{align}
where $\tilde{E} = E - t^2 / E$, $T= t + t^2 / E$. 
Apparently, Eq.~(\ref{ssha}) and (\ref{sshb}) together are reminiscent of the  eigenequations for the renowned Su-Schrieffer-Heeger (SSH) model~\cite{sshprl,ssh1980,PhysRevB.110.035106}.
As is well-known, the SSH model dwells in nontrivial topological phase with two degenerate zero-energy edge modes when the inter-cell hopping is greater then the intra-cell hopping.
Accordingly, one may reasonably infer that the trimer lattice model may demonstrate nontrivial topological properties when $T<J$ with $\tilde{E}=0$.
Back to the parameters of the original trimer model, one can predict that when \( J > 2 t \),  the trimer lattice model exhibits topological edge states with an energy $E=t$; additionally, when \( J > 0 \), the trimer lattice model possesses topological edge states with an energy $E=-t$.
This infer is clearly verified by numerical calculations as shown in Fig.~\ref{Fig02}, where we numerically calculate the energy spectrum of the trimer lattice model in Eq.~(\ref{Htrimer}) under open boundary condition (OBC) as a function of inter-cell hopping $J$ in the clean limit with $L=699$.

\begin{figure}[tbp]
\includegraphics[width=8.7cm]{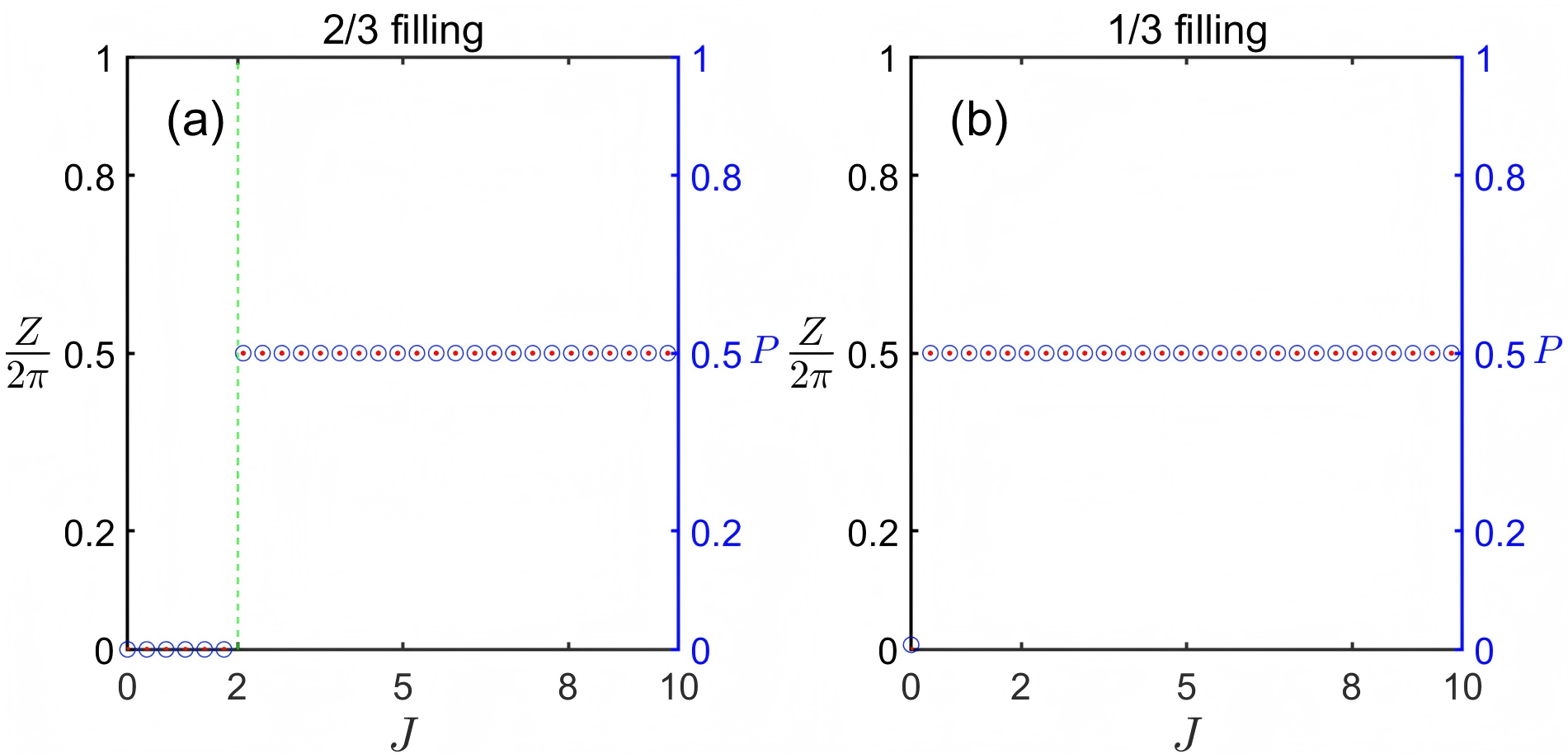}
\caption{\label{complexMME}
The total Zak phase $Z$ and polarization $P$ of the clean trimer lattice in Eq.~(\ref{Htrimer}) under different filling conditions.
(a) The two lower bands are filled.
(b) Only the lowest band is filled.
The red dots represent the Zak phase $Z$ and the blue circles represent polarization $P$. 
The Zak phase $Z$ is calculated in momentum space and 
the polarization $P$ is calculated in real space under periodic boundary condition (PBC) with $L=3N=699$. Common parameter $t=1$.
}\label{Fig03}
\end{figure}

On the other hand, the topological properties of the model in Eq.~(\ref{Htrimer}) under the clean limit can be characterized well by a topological invariant, namely the Zak phase~\cite{PhysRevLett.62.2747, PhysRevB.47.1651, RevModPhys.66.899, RevModPhys.84.1419, PhysRevB.92.041102, PhysRevLett.118.076803,PhysRevB.84.195452}. For a multi-band system, the Zak phase of the $n$-th band is defined as,
\begin{align}
Z_n = i \oint_{BZ} \langle u_{n,\mathbf{k}} \mid \partial_{\mathbf{k}} u_{n,\mathbf{k}} \rangle d\mathbf{k},  \label{zak}
\end{align}
where $u_{n,\mathbf{k}}$ is the Bloch eigenvector of the $n$-th band for the clean trimer lattice model and $\oint_{BZ}$ indicates the integral is taken around the loop formed by the 1D Brillouin zone. 
The emergence of topological edge states between $m$-th and ($m\!+\!1$)-th subbands is associated with the nontrivial total Zak phase $Z$ of $\pi$, with $Z=\sum_{n=1}^m Z_n$.
Fig.~\ref{Fig03} (left $y$-axis) shows the $J$-dependent total Zak phase $Z$ for the clean trimer lattice at two distinct fillings.
As shown in Fig.~\ref{Fig03}a, the total Zak phase $Z$ is nontrivial with value $\pi$ at $2/3$ filling  in the region $J>2$. 
Thus, the topological edge states emerge correpondingly in the energy gap between the second and third subbands in Fig.~\ref{Fig02}.
At $1/3$ filling, the total Zak phase $Z$ is nontrivial in the region $J>0$, see Fig.~\ref{Fig03}b. 
Therefore, there exist topological edge states in the region $J>0$ in Fig.~\ref{Fig02}.

As we will later focus on the topological properties of the most general case with quasiperiodic disorder, the Zak phase basically relying on Bloch wavefunctions will no longer be applicable. Therefore, one has to resort to alternative topological invariants. 
We adopt the polarization $P$ in the ground state based on the Resta's formula~\cite{PhysRevLett.80.1800,Resta2007,PhysRevLett.125.166801,Vanderbilt_2018},  
which is defined in the real space as,
\begin{align}
P = \frac{1}{2 \pi}  \text{Im}\ln \left[ \det(U^\dagger \hat{Q} U) \sqrt{\det(\hat{Q}^\dagger)} \right],   \label{polar} 
\end{align}
where $\hat{Q} \equiv \exp\left (  i2\pi  \hat{x}/{L}  \right )$
with $\hat{x}$ being the position operator along the lattice chain direction.
The matrix $U$ is constructed by column-wise packing all the occupied eigenstates according to different filling conditions, such that $U U^\dagger$ is the projector to the occupied subspace. 
Throughout this paper, $\ln$ denotes the natural logarithm and $\log$ denotes the logarithm with base $10$.
The polarization $P$ in Eq.~(\ref{polar}) is well defined and applicable to both correlated systems and independent-particle systems, as well as to crystalline and disordered systems~\cite{PhysRevLett.80.1800,PhysRevLett.125.166801}. 
In Fig.~\ref{Fig03} (right $y$-axis) , we show the polarization $P$ as a function of $J$ for the clean trimer lattice under two distinct fillings, 
which confirms that the system becomes topologically non-trivial in the region of $J> 2$ at $2/3$ filling,  while at $1/3$ filling, the system is topologically non-trivial in the region of $ J > 0$.

\begin{figure}[tbp]
\includegraphics[width=8.7cm]{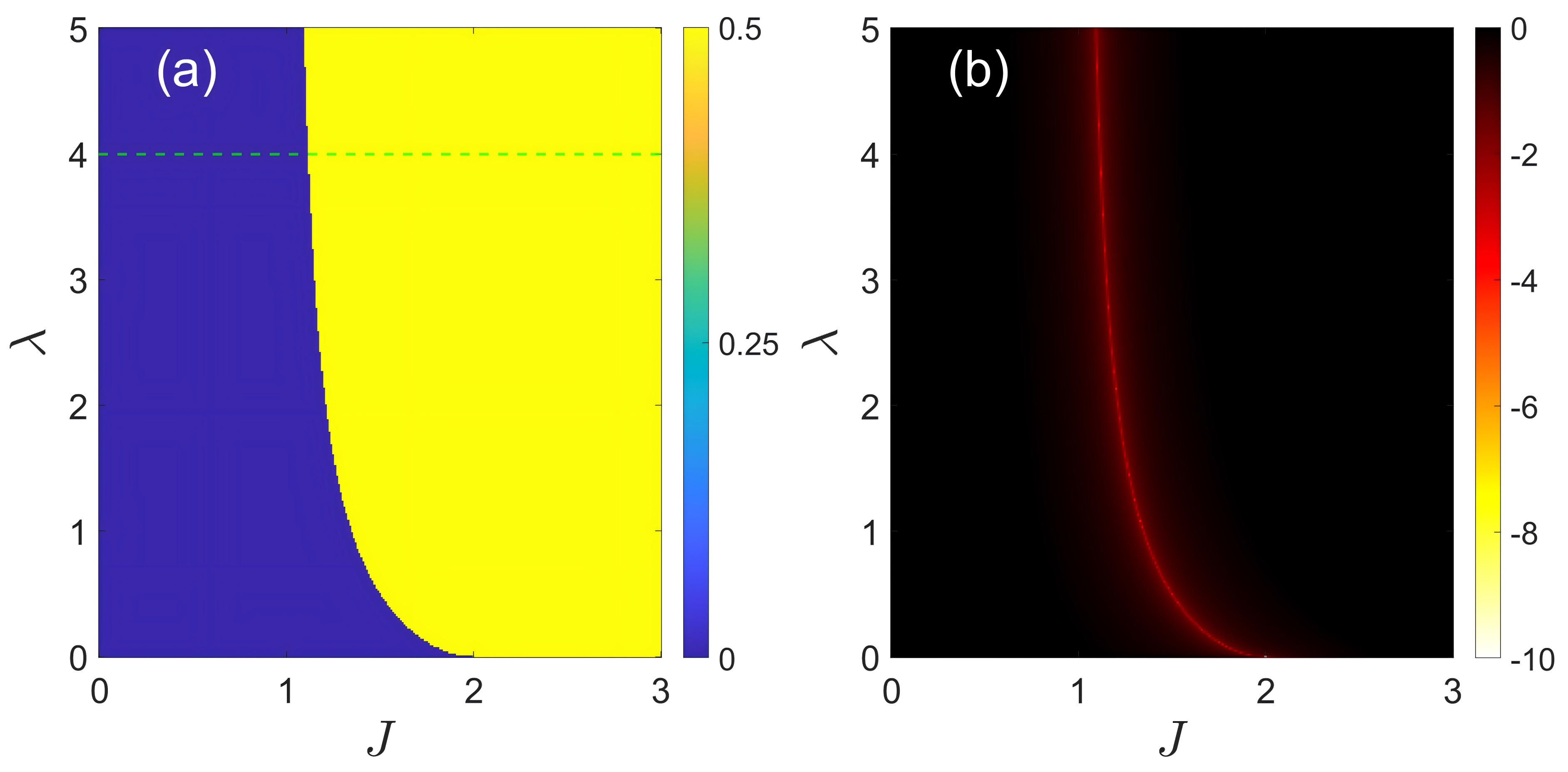}
\caption{
The topological phase diagram of the mosaic trimer lattice of Eq.~(\ref{Htrimer}) in the $\lambda-J$ plane at $2/3$ filling.
(a) The polarization $P$ as a function of $\lambda$ and $J$. Values of $P$ in the phase diagram are marked with different colors. The green dashed line corresponds to $\lambda=4$.
(b) The energy gap $\Delta E_{2/3}$  as a function of $\lambda$ and $J$. The color denotes the value of $\log \Delta E_{2/3} $.
Both the polarization $P$ and the energy gap $\Delta E_{2/3}$  are calculated under periodic boundary conditions (PBCs) with $L=3N=1830$. Common parameter $t=1$.
}\label{Fig04}
\end{figure}

\textcolor{blue}{\em Topological enhancement by quasiperiodicity}--With the preceding analysis on the clean limit, we now proceed to investigate the effect of quasiperiodic disorder on the existing topological phases in detail.
We first examine the fate of the topological phase at $2/3$ filling when subjected to the quasiperiodic mosaic modulations.
Once the quasiperiodic  disorder is introduced in, translational symmetry is broken and Bloch Hamiltonian is not available. 
Thus, we resort to the real-space topological invariant $P$ to characterize the topological phase. 
The corresponding results are shown in Fig.~\ref{Fig04}(a), where the polarization $P$ is calculated as a function of disorder strength $\lambda$ and the inter-cell hopping $J$.  When all the values of $P$ are denoted by color, a topological phase diagram in $\lambda-J$ plane is mapped out.
It is shown that the original topological phase at clean limit is dramatically enhanced as the strength of quasiperiodic disorder $\lambda$ increases. 
The quasiperiodic disorder renders part of the trivial phase into topological Anderson insulator (TAI).

Moreover, we also calculate the bulk energy gap $\Delta E_{2/3}$ for $2/3$ filling case under periodic boundary condition (PBC). 
The bulk energy gap is a quantity commonly used to detect phase transitions. Here we use it to confirm the phase boundary delineating the topological phase and the trivial phase. For the $2/3$ filling case, it is calculated as $\Delta E_{2/3}=E_{2N+1}-E_{2N}$. In Fig.~\ref{Fig04}(b), the bulk energy gap under PBC is plotted as a function of disorder strength $\lambda$ and the inter-cell hopping $J$. The color of each point stands for $\log \Delta E_{2/3}$.
Evidently, the topological phase boundary signaled by the bulk energy gap $\Delta E_{2/3}$ for $2/3$ filling agrees well with the one obtained by polarization $P$ as demonstrated in Fig.~\ref{Fig04}(a).

\begin{figure}[tbp]
\includegraphics[width=8.7cm]{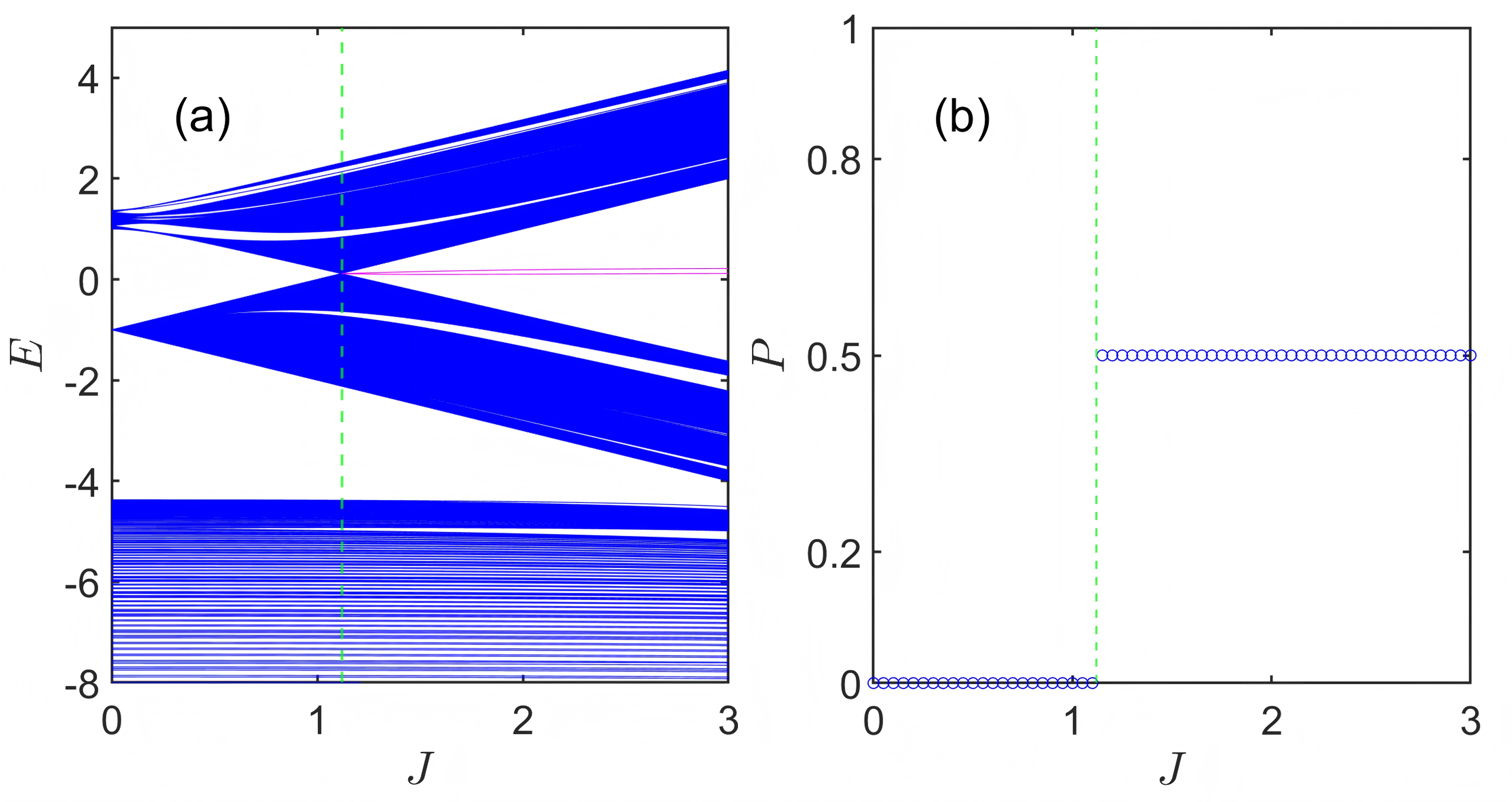}
\caption{
(a) The eigenenergy spectra $E$ around 2/3 filling as a function of $J$ with $\lambda=4$ and open boundary conditions (OBCs).
The emergence of topological edge states is highlighted with magenta lines.
(b) The polarization $P$ as a function of $J$  with $\lambda=4$. 
Both the eigenenergy spectrum $E$  and the polarization $P$ are calculated with $L=3N=1131$. Common parameter $t=1$.
The green dashed lines in (a) and (b) both correspond to \( J = 1.12 \), marking the topological phase transition point.
}\label{Fig05}
\end{figure}

To gain an intuitive visualization of the topological phase transition, we plot the  eigenenergy spectra $E$  of the upper two bands under open boundary condition (OBC) in Fig.~\ref{Fig05}(a). The strength of the quasiperiodic disorder is fixed at $\lambda=4$ as indicated by the green dashed line in Fig.~\ref{Fig04}(a).
Obviously, as highlighted by magenta color in Fig.~\ref{Fig05}(a), topological edge states emerge starting from $J=1.12$, indicating that the trimer lattice enters topological phase. 
This is in good agreement with the behavior of the  real space topological invariant $P$, as shown in Fig.~\ref{Fig05}(b).
For the regions on the two sides of $J=1.12$, the corresponding values of polarization $P$ are $0$ and $1/2$, respectively.
Both of the green dashed lines in Fig.~\ref{Fig05}(a) and Fig.~\ref{Fig05}(b) correspond to the same inter-cell hopping strength $J=1.12$.

Notably, one can observe that the energy of the topological edge states displayed in Fig.~\ref{Fig05}(a) is very close to zero.
This tendency of the topological edge states energy $E$ approaching zero can be further confirmed through numerical simulations with increasing quasiperiodic disorder strengths. 
Favorably, from this numerical observation, one can  analytically estimate the topological phase transition point at the limit of $\lambda \rightarrow \infty$.
For the most general case with quasiperiodic modulations $V_j$, one can also formally derive Eqs. (\ref{ssha}) and (\ref{sshb}) on the basis of the eigenequations Eqs.~(\ref{psia})-(\ref{psic}). The only difference is that the parameters $\tilde{E}$ and $T$ in Eqs. (\ref{ssha}) and (\ref{sshb}) take on new forms.
Specifically, here we have $\tilde{E} = E - t^2 / (E-V_j)$, $T= t + t^2 / (E-V_j)$. 
In the limit of $\lambda \rightarrow \infty$, $V_j\rightarrow \infty$. Combining the numerical observation $E\rightarrow 0$, we have $\tilde{E} =0$, $T= t$.
Therefore, in the limit of large quasiperiodic disorder strength, the topological phase transition for the $2/3$ filling case occurs at $J=t$.
This analytical deduction is consistent with the numerical results presented in both Fig.~\ref{Fig04}(a) and Fig.~\ref{Fig04}(b), where one can observe that the phase transition curve tends to approach $J=1$ as the quasiperiodic disorder strength $\lambda$ increases.

\begin{figure}[tbp]
\includegraphics[width=8.7cm]{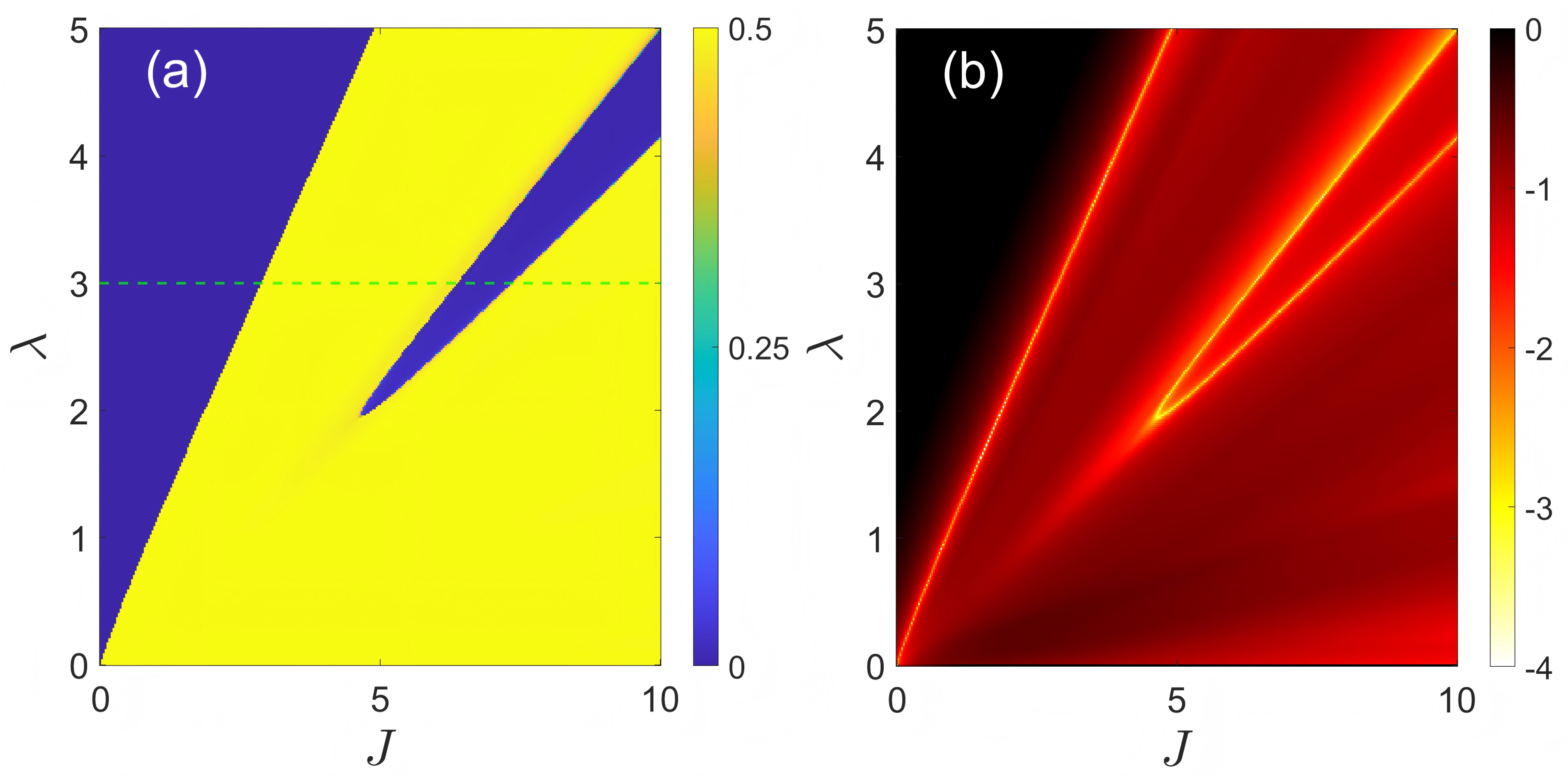}
\caption{
The topological phase diagram of the mosaic trimer lattice of Eq.~(\ref{Htrimer}) in the $\lambda-J$ plane at $1/3$ filling.
(a) The polarization $P$ as a function of $\lambda$ and $J$. Values of $P$ in the phase diagram are marked with different colors. The green dashed line corresponds to $\lambda=3$.
(b) The energy gap $\Delta E_{1/3}$ as a function of $\lambda$ and $J$. The color denotes the value of $\log \Delta E_{1/3} $.
Both the polarization $P$ and the energy gap $\Delta E_{1/3}$  are calculated under periodic boundary conditions (PBCs) for a trimer lattice with $L=3N=1830$. Common parameter $t=1$.
}\label{Fig06}
\end{figure}

\textcolor{blue}{\em Topological  suppression by quasiperiodicity and the reentrant topological phase transition}--In this section, we continue to investigate the effect of the quasiperiodic disorder on the other existing topological phase at $1/3$ filling.
In sharp contrast to the case discussed earlier, the quasiperiodic disorder exhibits a strong destructive effect on the topological phase of the clean limit.
As shown in Fig.~\ref{Fig06}(a), as the quasiperiodic disorder strength $\lambda$ increases, the region of the original topological phase shrinks significantly.
More interestingly, when the quasiperiodic disorder strength $\lambda$ grows strong enough, small island composed of trivial phase emerges inside the topological phase. This signals the emergence of the topological-trivial-topological reentrant topological phase transitions (RTPTs).
Notably, accordingly to the phase diagram in $\lambda-J$ plane demonstrated in Fig.~\ref{Fig06}(a), it can be seen that, not only the increasing of the quasiperiodic disorder strength $\lambda$ can induce reentrant topological phase transition, but also the increasing of the inter-cell hopping strength $J$ will lead to reentrant topological phase transition as indicated by the green dashed line.

Alternatively, to map out the phase diagram, we also calculate the bulk energy gap $\Delta E_{1/3}$ under periodic boundary condition for the $1/3$ filling case.
As shown in  Fig.~\ref{Fig06}(b),  by using the values of $\log \Delta E_{1/3}$ to color-code the plot, one can clearly identify the topological phase transition boundaries, which are in good agreement with the phase diagram presented in Fig.~\ref{Fig06}(a). 

\begin{figure}[tbp]
\includegraphics[width=8.7cm]{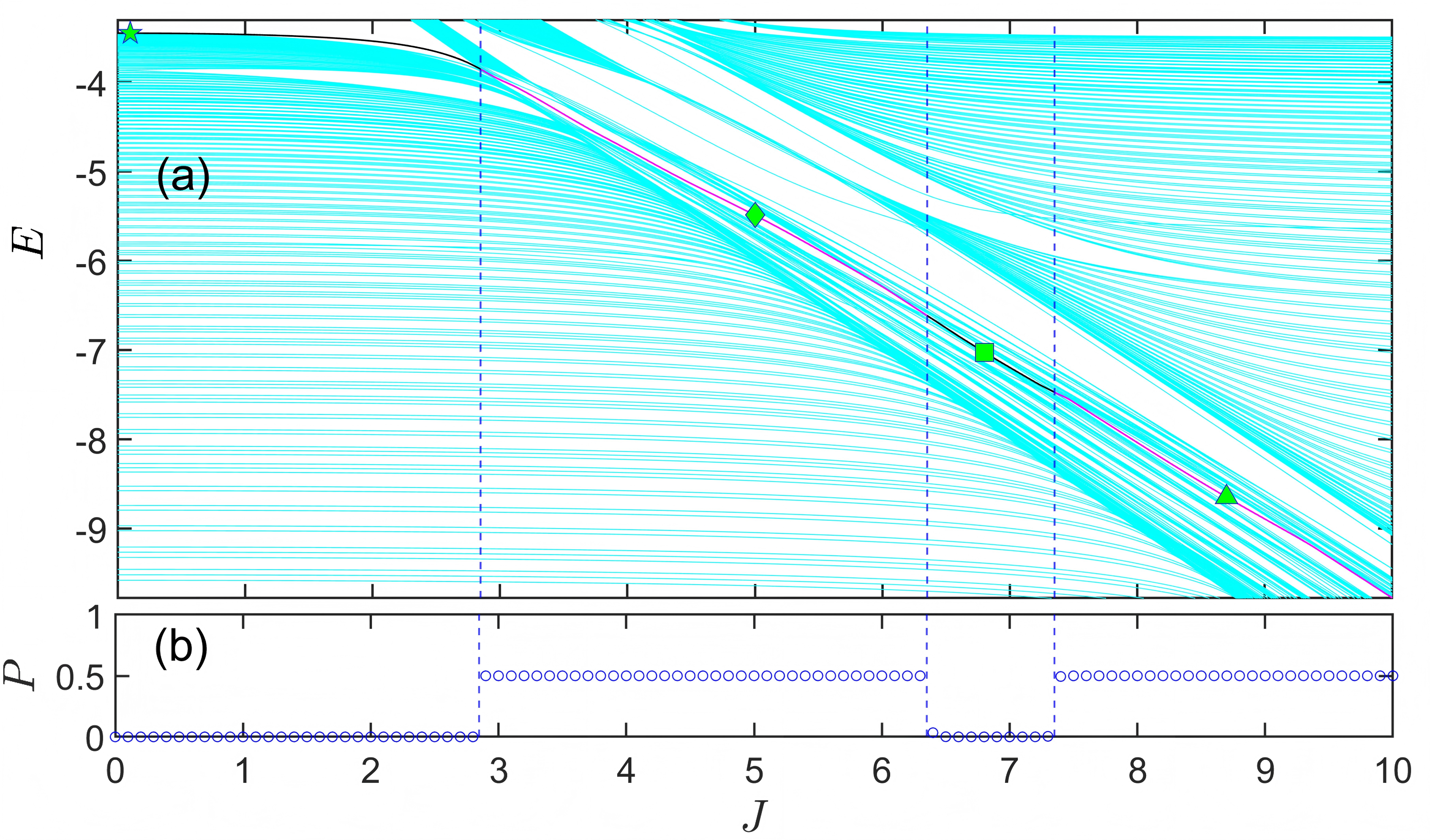}
\caption{
(a) The eigenenergy spectrum $E$ around 1/3 filling as a function of $J$ with $\lambda=3$ and open boundary conditions (OBCs).
The emergence of topological edge states is highlighted with magenta lines.
(b) The polarization $P$ as a function of $J$  with $\lambda=3$. 
Both the eigenenergy spectrum $E$  and the polarization $P$ are calculated with $L=3N=1131$. Common parameter $t=1$.
The blue dashed lines in both (a) and (b) mark the topological phase transition points.
}\label{Fig07}
\end{figure}

\begin{figure}[bp]
\includegraphics[width=7.7cm]{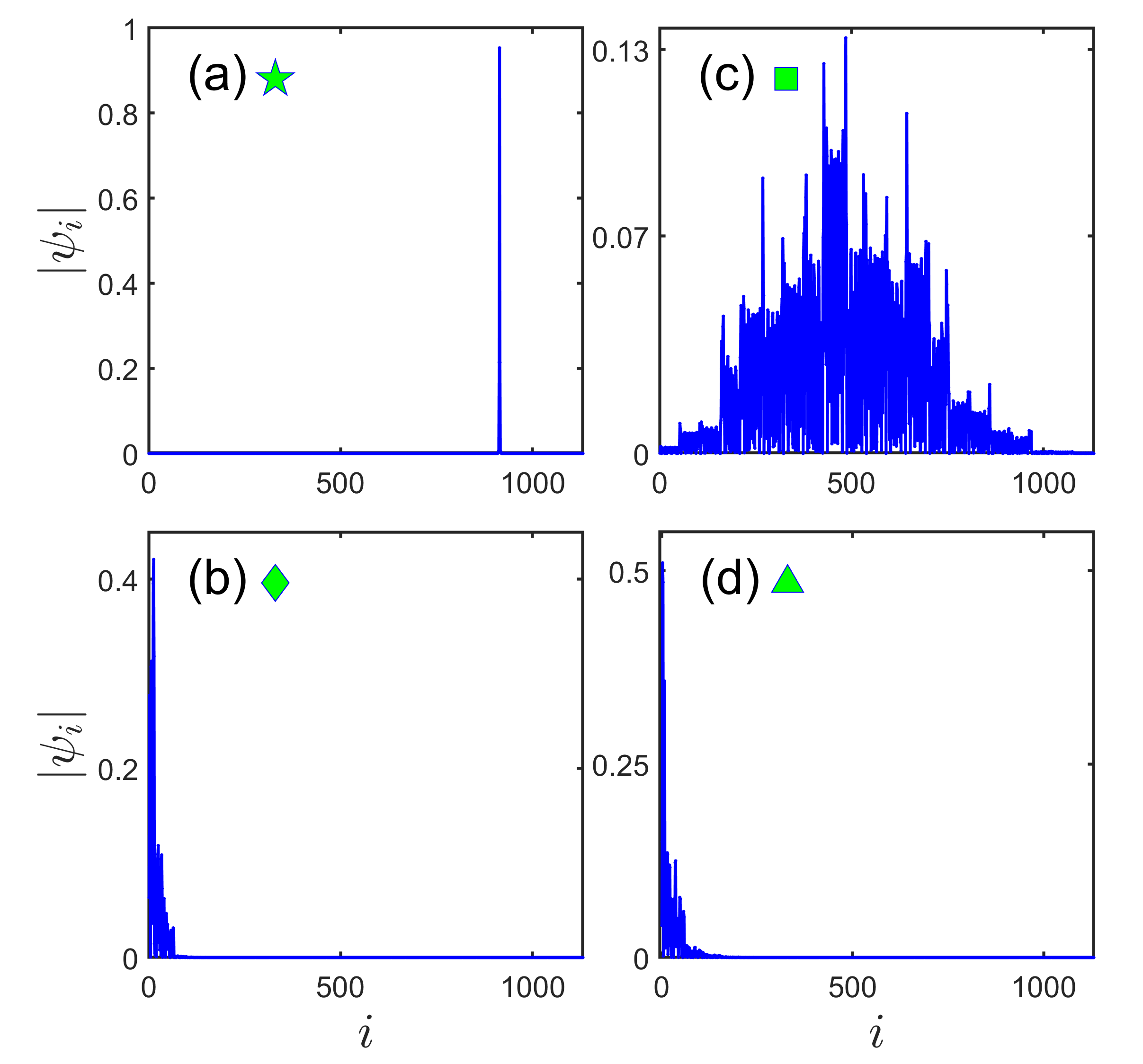}
\caption{
Typical probability amplitude distribution $\psi_i$ of the $N$-th eigenstate as a function of the lattice site $i$ for different phases. The quasiperiodic disorder is fixed at $\lambda=3$. The values of  inter-cell hopping $J$ are highlighted by four different markers as shown in Fig. \ref{Fig07}.
(a) $J=0.1$.
(b) $J=5$.
(c) $J=6.8$. 
(d) $J=8.7$.
Lattice size is $L=3N=1131$ and $t=1$.
}\label{Fig08}
\end{figure}

To showcase the emergent reentrant topological phase transitions occurring in the $\lambda-J$ plane , we present in Fig.~\ref{Fig07}(a) the energy spectra around $1/3$ filling under open boundary condition as a function of the inter-cell hopping strength $J$ at fixed quasiperiodic disorder strength $\lambda=3$.
Correspondingly, the real space topological invariant $P$ is calculated as a function of the inter-cell hopping $J$ to identify the topological phases, as shown in Fig.~\ref{Fig07}(b).
Obviously, owing to the disruptive effect of quasiperiodic disorder, the trimer lattice first dwells in a trivial phase. 
Subsequently, with the increase of the inter-cell hopping $J$, the system undergoes an interesting reentrant topological phase transition featuring the topological-trivial-topological sequence.
Owing to the dense distribution of energy levels, for the sake of visual identification, we have highlighted the $N$-th energy level and marked its topological region in magenta in Fig.~\ref{Fig07}(a). 
To visualize the reentrant topological phase transitions more clearly, 
in Fig.~\ref{Fig08} we present the probability amplitude distribution profiles of several typical eigenstates, which correspond to parameter points from different phases, as denoted by distinct markers in Fig.~\ref{Fig07}(a).
As shown in Fig.~\ref{Fig08}(a), the trimer lattice is in a trivial phase. Since the inter-cell hopping $J$ is much smaller than the quasiperiodic disorder $\lambda$, the $N$-th eigenstate turns out to be a normal Anderson localized state.
As the inter-cell hopping $J$ increases, the system enters a topological phase, and topological edge state emerges, as illustrated in Fig.~\ref{Fig08}(b).
When the inter-cell hopping $J$ continues to increase, the trimer lattice enters a trivial phase again, and the $N$-th eigenstate evolve into a ordinary quantum states, as depicted in Fig.~\ref{Fig08} (c).
If one increases the inter-cell hopping $J$ further, the systems re-enters another topological phase. Correspondingly, the eigenstate turns into topological edge state, which can be seen clearly in Fig.~\ref{Fig08} (d).

\begin{figure}[tbp]   
\includegraphics[width=8.7cm]{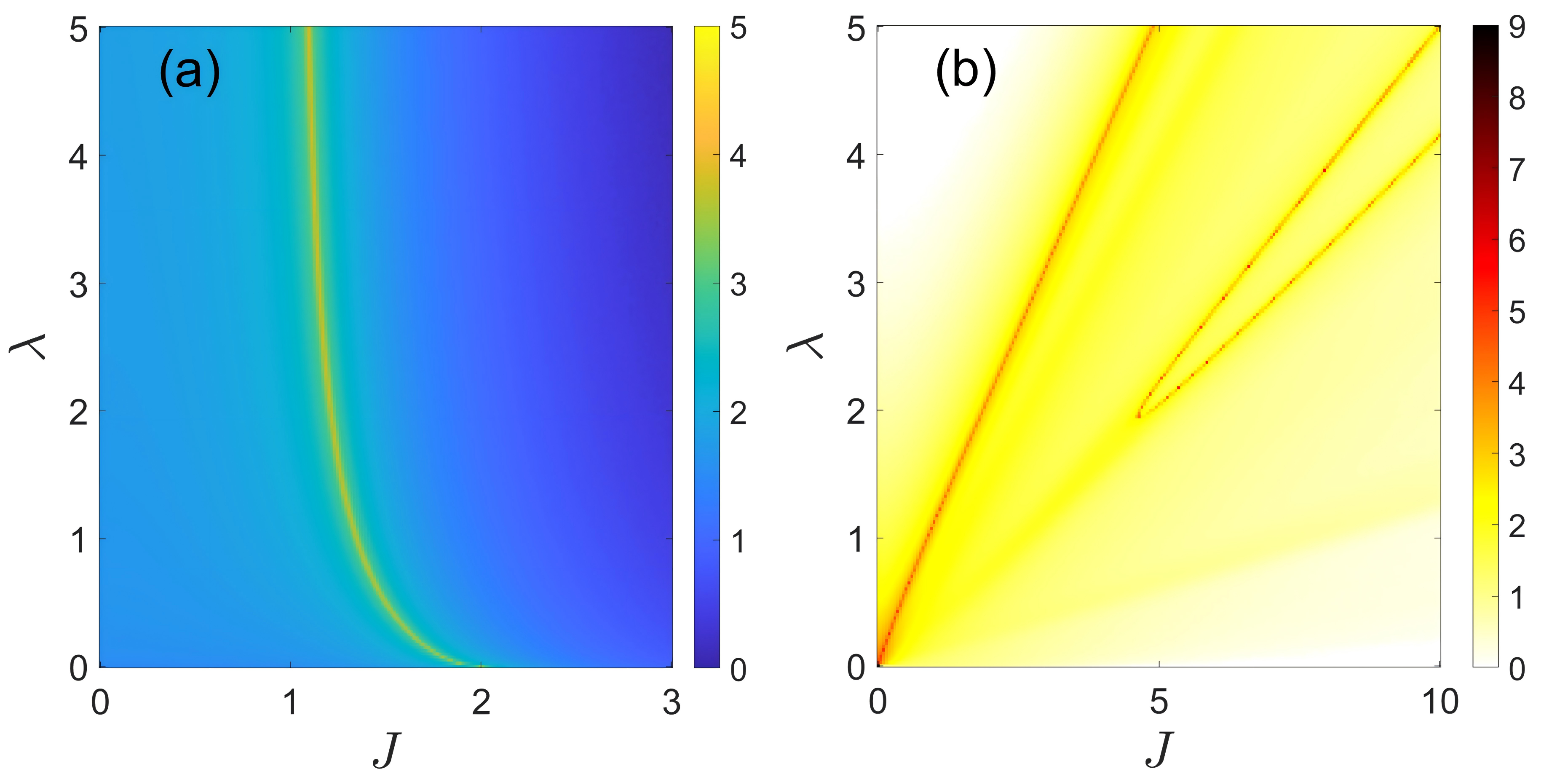}
\caption{The topological phase diagrams spanned in $\lambda-J$ plane detected by the many-body ground state fidelity susceptibility $\chi_{gs}$ under periodic boundary conditions.
(a) $2/3$ filling.
(b) $1/3$ filling.
The color of each point ($\lambda, J$) is encoded by $\log \chi_{gs}$.
The minor perturbation to $J$ is chosen as $\delta J = 10^{-6}$.
The lattice size is $L=3N=1830$. Common parameter $t=1$.
}\label{Fig09}
\end{figure}

\textcolor{blue}{\em Fidelity susceptibility}--In this section, we utilize the many-body ground state fidelity susceptibility~\cite{PhysRevLett.96.140604,PhysRevE.74.031123,PhysRevE.76.022101,PhysRevE.76.061108,PhysRevA.77.032111,FidelityGu,PhysRevA.98.052116,Sun2021, PhysRevResearch.3.013015,Tu2023generalpropertiesof} to further characterize the topological phase transitions of the quasiperiodic disorder modulated trimer lattice. As a well established concept in quantum many-body physics, fidelity susceptibility measures how sensitive a quantum system is when subjected to small external perturbations and proves to be a powerful tool for identifying quantum phase transitions.
The ground-state fidelity susceptibility $\chi_{gs}$ can be calculated as~\cite{PhysRevA.77.032111,FidelityGu},
\begin{align}
\chi_{gs} = -\frac{\partial^2 F_{gs}}{\partial (\delta J)^2} = \lim_{\delta J \to 0} \frac{-2 \ln F_{gs}}{(\delta J)^2},
\end{align}
where $F_{gs}$ is the many-body ground state fidelity~\cite{PhysRevLett.96.140604,PhysRevE.74.031123,PhysRevE.76.022101,PhysRevE.76.061108} defined as
$F_{gs} =\vert \langle G (J+\delta J) \vert G (J)\rangle\vert$ with $\vert G (J)\rangle$ being the system's ground state wavefunction under certain fillings with inter-cell hopping $J$.
In practical numerical calculations, $\delta J$ is a sufficiently small variation. Here, we take $\delta J=10^{-6}$.

As shown in Fig.~\ref{Fig09}, we calculate the many-body ground state fidelity susceptibility $\chi_{gs}$ for the quasiperiodic disorder modulated trimer lattice at $2/3$ filling (Fig.~\ref{Fig09}a) and $1/3$ filling (Fig.~\ref{Fig09}b), respectively. 
In the $\lambda-J$ plane, the color of each data point corresponds to the value of $\log \chi_{gs}$. 
The peaks of many-body ground state fidelity susceptibility $\chi_{gs}$ clearly signify the phase boundaries of topological phase transitions, which are in good agreement with Fig.~\ref{Fig04} and Fig.~\ref{Fig06}, respectively.

\textcolor{blue}{\em Summary}--In this work, we explore the specific and rich roles that quasiperiodic disorder plays on topological properties. 
To this end, a one-dimensional quasiperiodic-potential-modulated mosaic trimer lattice is constructed.
Analytical analysis shows that, in clean limit free of quasiperiodic disorder the trimer lattice exhibits nice topological properties, possessing two topological phases corresponding to the $2/3$ filling case and $1/3$ filling case respectively. 
Taking this fact as the starting point, we examined the topological effects of quasiperiodic disorder.
It turns out the topological roles of the quasiperiodic disorder is multiple in this single model.
For the topological phase at $2/3$ filling,  the quasiperiodic disorder exerts an evident topological enhancement effect. 
Some regions originally belonging to the trivial phase are transformed into topologically non-trivial regions. 
Correspondingly, the topological Anderson insulator (TAI) phase emerges.
On the other hand, for the topological phase at $1/3$ filling,  the quasperiodic disorder exhibits a completely opposite effect. 
On the whole, the introduction of quasiperiodic disorder can significantly impair the topological phase of the trimer lattice, with the topological nontrivial regions being obviously reduced. 
More interestingly, quasiperiodic disorder can induce small island composed of the topological trivial phase within the original topological phase.
This key fact implies that in the mosaic trimer lattice, not only can the quasiperiodic disorder strength $\lambda$ induce reentrant topological phase transitions, but also the inter-cell hopping strength $J$ can drive the system to undergo reentrant topological phase transitions.
The topological phase are characterized by polarization $P$, a real space topological invariant, and confirmed by demonstrations of relevant eigenenergy spectra and wavefunction distributions.
Furthermore, the topological phase diagrams at different fillings are  verified by numerical computations of bulk energy gap and the many-body ground state fidelity susceptibility. 
This study will further facilitate the research on the diverse and novel roles of quasiperiodic disorder, besides localization effects and topological actions.

\textcolor{blue}{\em Acknowledgments}-- L. W. is supported by the Research Project Supported by Shanxi Scholarship Council of China (Grant No. 2024-004), the Fundamental Research Program of Shanxi Province, China (Grant No. 202203021211315), the National Natural Science Foundation of China (Grant Nos. 11404199, 12147215) and the Fundamental Research Program of Shanxi Province, China (Grant Nos. 1331KSC and 2015021012).
S. C. is supported by National Key Research and Development Program of China (Grant No. 2021YFA1402104) and the NSFC under Grants No. 12474287, No. 12547107, and No. T2121001.

\end{document}